# Fabrication and Properties of NbN/NbN$_x$/NbN and Nb/NbN$_x$/Nb Josephson Junctions

Sergey K. Tolpygo, *Senior Member, IEEE*, Ravi Rastogi, David Kim, Terence J. Weir, Neel Parmar, and Evan B. Golden

*Abstract*—Increasing integration scale of superconductor electronics (SCE) requires employing kinetic inductors and self-shunted Josephson junctions (JJs) for miniaturizing inductors and JJs. We have been developing a ten-superconductor-layer planarized fabrication process with NbN kinetic inductors and searching for suitable self-shunted JJs to potentially replace high Josephson critical current density, $J_c$, Nb/Al-AlO$_x$/Nb junctions. We report on the fabrication and electrical properties of NbN/NbN$_x$/NbN junctions produced by reactive sputtering in Ar+N$_2$ mixture on 200-mm wafers at 200 ºC and incorporated into a planarized process with two Nb ground planes and Nb wiring layer. Here NbN is a stoichiometric nitride with superconducting critical temperature $T_c$ =15 K and NbN$_x$ is a high resistivity, nonsuperconducting nitride deposited using a higher nitrogen partial pressure than for the NbN electrodes. For comparison, we co-fabricated Nb/NbN$_x$/Nb JJs using the same NbN$_x$ barriers deposited at 20 ºC. We varied the NbN$_x$ barrier thickness from 5 nm to 20 nm, resulting in the range of $J_c$ from about 1 mA/µm$^2$ down to ~10 µA/µm$^2$, and extracted coherence length of 3 nm and 4 nm in NbN$_x$ deposited, respectively at 20 ºC and 200 ºC. Both types of JJs are well described by resistively and capacitively shunted junction model without any excess current. We found the $J_c$ of NbN/NbN$_x$/NbN JJs to be somewhat lower than of Nb/NbN$_x$/Nb JJs with the same barrier thickness, despite a much higher $T_c$ and energy gap of NbN than of Nb electrodes. $I_cR_n$ products up to ~ 0.5 mV were obtained for JJs with $J_c$~ 0.6 mA/µm$^2$. $J_c(T)$ dependences have been measured.

## I. INTRODUCTION

INCREASING integration scale of superconductor electronics (SCE) toward levels needed for advanced computations requires employing kinetic inductors instead of geometrical ones and self-shunted Josephson junctions (JJs) instead of externally shunted ones in order to enable miniaturization of these crucial components of superconductor integrated circuits [1], [2], [3]. During the last several years, we have been developing at MIT Lincoln Laboratory (MIT LL) a ten-superconductor-layer process incorporating NbN and bilayer NbN/Nb kinetic inductors above the layer of Josephson junctions, the SFQ7ee process [3], [4]. At the same time, we have been searching for suitable self-shunted Josephson junctions to replaced externally shunted Nb/Al-AlO$_x$/Nb JJs which are otherwise hysteretic at all $J_c$ values used in the existing MIT LL process nodes up to about 0.6 mA/µm$^2$. The self-shunted JJs should satisfy at least three requirements in addition to the scalability requirement:

a) have sufficiently high $I_cR_n$ products to support high clock rates which make SCE digital circuits, along with ultralow energy dissipation, competitive with semiconductor CMOS circuits; $I_c$ is the junction critical current and $R_n$ is the internal damping resistance.

b) The minimum $I_c$ of the JJs, $I_{c,min}$ should be small enough to provide the ultralow energy dissipation and be limited from below only by the acceptable bit error rate of the circuits, e.g., $I_{c,min}$~ 50 µA. This sets the minimum required Josephson current density $j_{c,min} = I_{c,min}/A_{J,min}$, where $A_{J,min}$ is the minimum fabrication-process-defined junction area.

c) The most difficult requirement comes from the need to match impedance of the JJs and the wave impedance, $Z_0$ of passive transmission lines (PTLs) providing on-chip and chip-to-chip data transfer. For instance, the standard superconducting striplines in the SFQ5ee fabrication process at MIT LL have $Z_0 \approx$ 10 Ω at $w =$ 2 µm [5], where $w$ is the stripline width. To enable a very large-scale integration (VLSI), it is highly desirable the PTL width to be in the micrometer or even sub-micrometer range. Since $Z_0$ scales approximately as $1/w$, impedance matching requires JJs with high $R_n$ and $R_nA_J$ product.

These requirements are relatively easy to satisfy using JJs with medium-to-high transparency tunnel barriers, having $I_cR_n$~1 mV at $j_c$~1 mA/µm$^2$ [6]. In this case, $A_{J,min}$~0.05 µm$^2$ is provided by circular JJs with diameter $d_{min}$~ 0.25 µm, having $R_{n,max} = I_cR_n/I_{c,min}$ ~ 20 Ω that matches $Z_0$ of striplines with $w \approx$ 0.8 µm; see [5, Fig. 4].

On the other hand, these requirements are practically impossible to satisfy using SNS-type JJs, where N is a normal metal, or using highly disordered, high resistivity materials and doped semiconductors with low carrier concertation and resonant tunneling type conduction as the N-barrier; see, e.g.

This material is based upon work supported by the Under Secretary of War for Research and Engineering under U.S. Air Force Contract No. FA8702-15-D-0001. *(Corresponding author: Sergey K. Tolpygo).*

All the authors are with Lincoln Laboratory, Massachusetts Institute of Technology, Lexington, MA 02421, USA (emails: sergey.tolpygo@ll.mit.edu; ravi.rastogi@ll.mit.edu; david.kim@ll.mit.edu; weir@ll.mit.edu; neel.parmar@ll.mit.edu; evan.golden@ll.mit.edu).

Color versions of one or more of the figures in this article are available online at http://ieeexplore.ieee.org

[7], [8], [9], [10], [11], [12], [13], [14], [15]. In the sandwich- and bridge-type JJs, the only adjustable parameters are the barrier thickness (bridge length), $t$ and the barrier (bridge) material resistivity, $\rho_n$. The junction resistance $R_n A_J$ is proportional to $t$ while $j_c$ exponentially decreases with increasing $t$ as $\exp(-t/\xi_n)$, where $\xi_n$ is the coherence length in the barrier, which also decreases with increasing the barrier resistivity as approximately $\rho_n^{-1/2}$. In the high resistivity barriers, $\xi_n$ becomes very short, a few nanometers. Therefore, it is practically impossible to achieve simultaneously the required high values of $R_n$, $j_c$, and $I_c R_n$ in the same junction.

To escape from the described conundrum, the authors of [16] proposed a bridge-type JJ with an additional adjustment relying on the contact resistance, $R_c$ between the dissimilar N and S materials in their overlap area that can be changed independently of the N-bridge cross section. The total junction resistance $R_n = 2R_c + R_{bridge}$ could then be made dominated by the $R_c$ while the critical current mainly determined by the bridge length and $\xi_n$. Although interesting theoretically, such a JJ structure does not appear to be more practical, scalable or reproducible that the regular trilayer tunnel barrier junctions [17] from the fabrication standpoint. It requires a highly reproducible fabrication of more features than a tunnel JJ: a deep submicron-scale planar bridge, and two high-contact-resistance interfaces between the N and S materials with tightly controlled overlap area, which are the analog of two sandwich junctions.

In the SFQ7ee process node, JJs are directly contacted by the NbN kinetic inductance layer; see [3, Fig. 12] and [15, Fig. 1]. To eliminate any potential junction degradation caused by diffusion of nitrogen into the JJ top (counter) electrode during NbN deposition at elevated temperatures or other causes related to processing dissimilar wiring layer and counter electrode metals, we investigated Josephson junctions with NbN electrodes. Because of a limited number of deposition chambers and sputtering targets in our PVD deposition cluster, we utilized a nonsuperconducting, high resistivity $NbN_x$ as a barrier material [15] in $NbN/NbN_x/NbN$ trilayer junctions. For comparison, we fabricated in the same process run also $Nb/NbN_x/Nb$ trilayer junctions.

## II. Fabrication of $NbN/NbN_x/NbN$ and $Nb/NbN_x/Nb$ Trilayer Junctions

All trilayers were deposited in an Endura PVD (Applied Materials, Inc.) cluster using magnetron sputtering on 200 mm oxidized Si wafers containing a patterned 200-nm-thick Nb layer (layer M4 in our nomenclature) covered by a 200-nm-thick planarized layer of $SiO_2$ interlayer dielectric. The deposition parameters for Nb, NbN, and $NbN_x$ films are given in Table I. The thickness of the bottom and top electrodes of the JJs was 150 nm for both NbN and Nb electrodes. The resistivity data in Table I were calculated using the average of the sheet resistance measured in 49 points on the witness films deposited on $SiO_2$-coated 200-mm wafers. The $SiO_2$ film was deposited using Plasma-Enhanced Chemical Vapor Deposition (PECVD).

TABLE I
Deposition Parameters of Nb, NbN, and $NbN_x$ Films

| Material and junction type | Power (kW) | Pressure (mT) | Ar flow (sccm) | $N_2$ flow (sccm) | $N_2$ partial pressure, $N_2/(N_2+Ar)$ | Deposition temperature (°C) | Film thickness (nm) | $\rho_{300K}$ (μΩ cm) | RRR = $R_{300}/R_{T>T_c}$ |
|---|---|---|---|---|---|---|---|---|---|
| Nb film | 1.2 | 4 | 40 | N/A | N/A | 20 | 200 | 19.6 | 4.7 |
| NbN film | 1.5 | | 60 | 20 | 0.25 | 200 | 200 | 190.6 | ~1 |
| $NbN_x$ barrier film for $Nb/NbN_x/Nb$ | 1.5 | | 10 | 95 | 0.905 | 20 | 50 | 925[a] | 0.47 |
| $NbN_x$ film for $NbN/NbN_x/NbN$ | 1.0 | | 10 | 150 | 0.938 | 200 | 20 | 1018[b] | 0.586 |

[a] Measured on a film deposited on PECVD $SiO_2$ on Si wafer
[b] Measured on a film deposited on PECVD $SiO_2$ on Si wafer

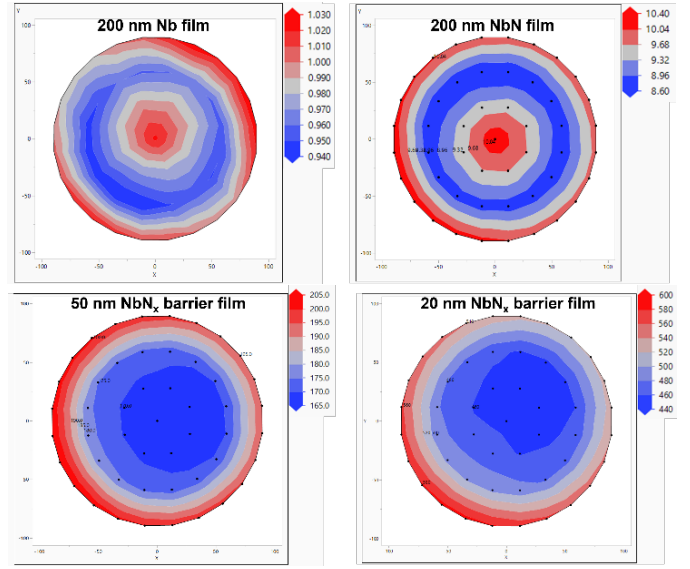

Fig. 1. Sheet resistance wafermaps at room temperature for the typical Nb, NbN, and $NbN_x$ films used in the fabrication of Josephson junctions. Deposition parameters for these films are given in Table I.

The typical sheet resistance wafermaps for the Nb and NbN films used for junction electrodes are shown in Fig. 1 (top row). These wafermaps have nearly cylindrical symmetry with the radial variation reflecting the sputtering rate pattern of a circular magnetron. The films are thinner near the rim and in the center of the wafers and thicker in the middle. The mean values are 0.987 Ω/sq and 9.54 Ω/sq for Nb and NbN, respectively, with standard deviations, $1\sigma_R$ of 0.029 Ω/sq and 0.635 Ω/sq. The total variation range, from the minimum to the maximum value, for the Nb film is from 0.94 to 1.03 Ω/sq and from 8.56 to 10.35 Ω/sq for the NbN film. Both ranges closely correspond to $3\sigma_R$. The relative variation $1\sigma_R/R_{mean}$ increases from about 3% for Nb to 6.7% for NbN films. This is a result of the sputtering gas mixture delivery and distribution in the Endura PVD sputtering chambers which, strictly speaking, are not designed and optimized for reactive sputtering.

It is well known that increasing nitrogen content in $Ar+N_2$ sputtering gas mixture above the optimum composition for producing NbN film with $T_c$ in the range from 14 to 16 K increases the film resistivity and eventually results in high-resistivity, nonsuperconducting $NbN_x$ (x>1) films [4], [15].

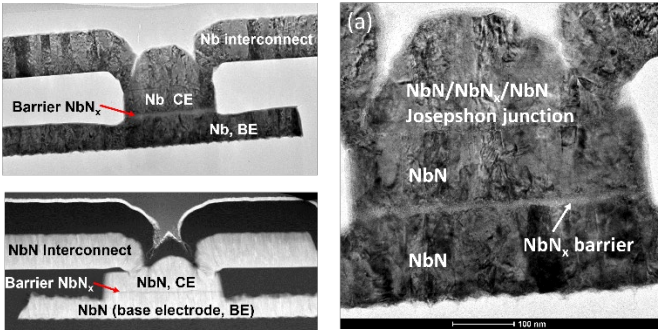

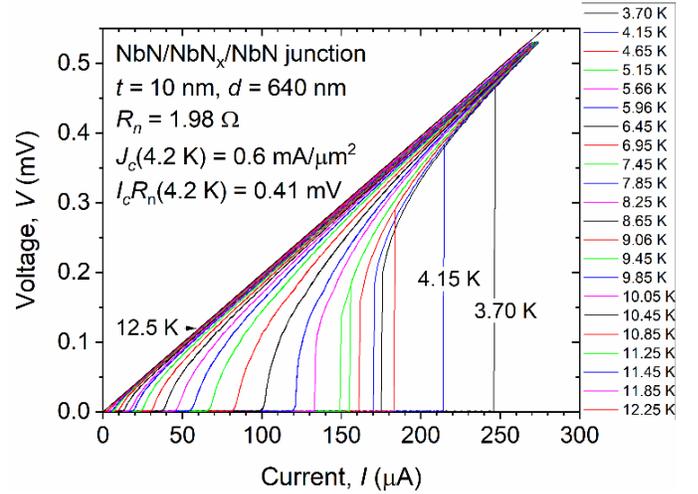

Fig. 2. Scanning Transmission Electron Microscope (STEM) images of the cross sections of Nb/NbN$_x$/Nb (left, top) and NbN/NbN$_x$/NbN (left bottom and right) junctions. A top via to the junction's counter electrode (CE) was etched in the SiO$_2$ dielectric planarizing the junction mesa, and either Nb or NbN interconnect layer was deposited and patterned.

Fig. 3. I-V characteristics of a NbN/NbN$_x$/NbN junctions with $t$ =10 nm NbN$_x$ barrier at different temperatures from 3.70 K to 12.25 K, from right to left in the graph. The critical temperature of NbN electrodes is $T_c$ =14.8 K. Above ~ 9 K, the I-V characteristics become noticeably affected by thermal noise.

Because of a limited number of different sputtering targets that can be accommodated in the same PVD cluster, we decided to use the high-resistivity NbN$_x$ films as the junction barrier [15].

We used slightly different parameters for depositing the barrier in NbN/NbN$_x$/Nb and in Nb/NbN$_x$/Nb trilayer junctions. NbN films with $T_c \approx 15$ K require deposition at 200 °C and, hence, the NbN$_x$ barrier films were deposited at the same temperature. In contrast, our Nb films are deposited at room temperature. To prevent their potential degradation at the interface with NbN$_x$ due to nitrogen diffusion at elevated temperatures, the NbN$_x$ barrier on Nb electrodes was deposited also at room temperature. Consequently, the sputtering power and gas mixture were adjusted to obtain nearly the same NbN$_x$ film resistivity as for the NbN$_x$ films deposited at 200 °C; see Table I.

The barrier film thickness was varied from 5 nm to 20 nm by adjusting the deposition time. The typical sheet resistance wafermaps for the NbN$_x$ films are shown in Fig. 1 (bottom row). These wafermaps are not as symmetrical as the stoichiometric NbN films and also show larger sheet resistance variations. For the 50-nm NbN$_x$ film deposited at 20 °C: $R_{mean}$ =185.0 Ω/sq, $1\sigma_R$ =14.7 Ω/sq, $1\sigma_R/R_{mean}$ = 7.9%, and the min-to-max range is from 166 to 206 Ω/sq. For the 20-nm NbN$_x$ film deposited at 200 °C: $R_{mean}$ =509.5 Ω/sq, $1\sigma_R$ =46.7 Ω/sq, $1\sigma_R/R_{mean}$ = 9.2%, and the min-to-max range is from 447.8 to 587.3 Ω/sq or $3\sigma_R$.

The described nonuniformities are caused mainly by the film thickness and nitrogen content nonuniformities. The latter slightly increases from the center toward the edges of wafer because of the way N$_2$ and Ar are delivered and distributed in the Endura PVD chamber. It is possible that the thickness nonuniformity increases even more with decreasing the average film thickness but we have not investigated this issue. The barrier thickness uniformity directly affects the uniformity of the Josephson critical current density on the wafers, $j_c$ which will be presented in the next section. From the sheet resistance data in Fig. 1, the $j_c$ uniformity is not expected to be great and expected to be worse than for Nb/Al-AlO$_x$/Nb tunnel junctions having the min-to-max $j_c$ variation of less than 10% [18], [19].

STEM images of the fabricated junctions are shown in Fig. 2. The junction fabrication is similar to the fabrication of Nb/Al-AlO$_x$/Nb junctions in [3], [20], except that we used shallow top vias to the junctions' counter (top) electrode (CE), which were etched in the SiO$_2$ dielectric planarizing the junction mesa after pattering the CE and the base (bottom) electrode (BE) of the junctions. Junction anodization processed used to protect the tunnel barrier in Nb/Al-AlO$_x$/Nb junctions [20] was not used for NbN$_x$-barrier junctions. Optical emission end-point detection and timed overetching was used to stop the etching on the base electrode after etching through the NbN$_x$ barrier. This overetching roughens the top surface of the remaining NbN base electrode outside the junctions as visible in the cross section in the left part of Fig. 2.

### III. ELECTRICAL TESTING RESULTS

#### A. Current-Voltage (I-V) Characteristics

Current-voltage characteristics (CVCs) of the junctions were measured using both a liquid helium immersion probe and a closed cycle cryocooler. The typical I-V characteristics of a circular NbN/NbN$_x$/Nb junction with diameter $d$ =680 nm and the barrier thickness $t$ =10 nm at various temperatures from 3.70 K to 12.25 K are shown in Fig. 3. At low temperatures, the junctions with this and lower barrier thicknesses are hysteretic with the Stewart-McCumber parameter $\beta_c = 2\pi I_c R_n^2 C/\Phi_0 > 1$, where $C$ is the junction capacitance and $\Phi_0$ the flux quantum. Hysteresis quickly disappears with increasing temperature because the critical current decreases and $\beta_c$ becomes less than 1; see below.

CVCs of the junctions with thicker barriers are nonhysteretic at 4.2 K as shown in Fig. 4 for NbN/NbN$_x$/NbN junctions with $t$ =15 and 20 nm. The junctions are well described by the RCSJ model without any excess current. CVCs of all junctions with $t$ = 20 nm can be fitted very well by the resistively shunted junction (RSJ) model $V = V_c(i^2 - 1)^{1/2}$, where $V_c = I_c R_n$ and $i = I/I_c$, neglecting the junction capacitance and indicating that $\beta_c \lesssim 0.1$; see solid red curve in Fig. 4.

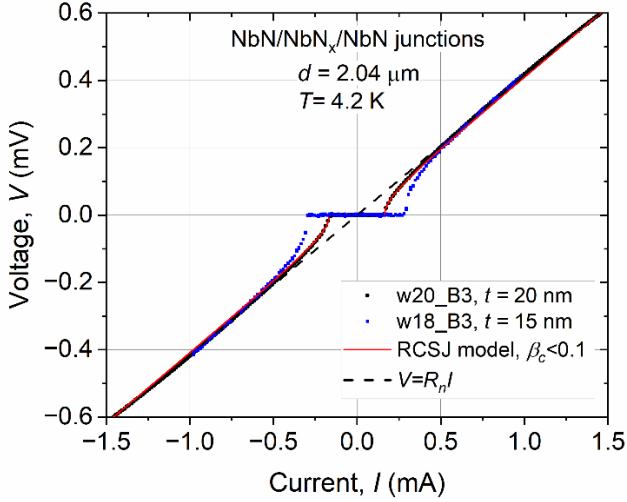

Fig. 4. *I-V* characteristics of the circular NbN/NbN$_x$/NbN junctions with $t$ =15 nm (wafer w18, location B3) and $t$ =20 nm (wafer w20, location B3) NbN$_x$ barriers at LHe temperature. At this barrier thickness, the junctions are nonhysteretic and are well described by the RSCJ model without any excess current.

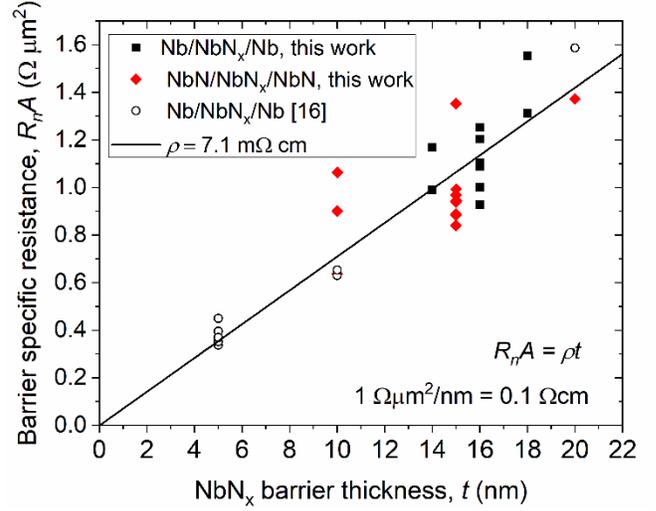

Fig. 6. Dependence of the junction specific resistance $R_nA$ of Nb/NbN$_x$/Nb and NbN/NbN$_x$/NbN Josephson junctions on the thickness of the NbN$_x$ barriers. The data for Nb/NbN$_x$/Nb junctions from [16], fabricated with the same NbN$_x$ deposition parameters, were also included.

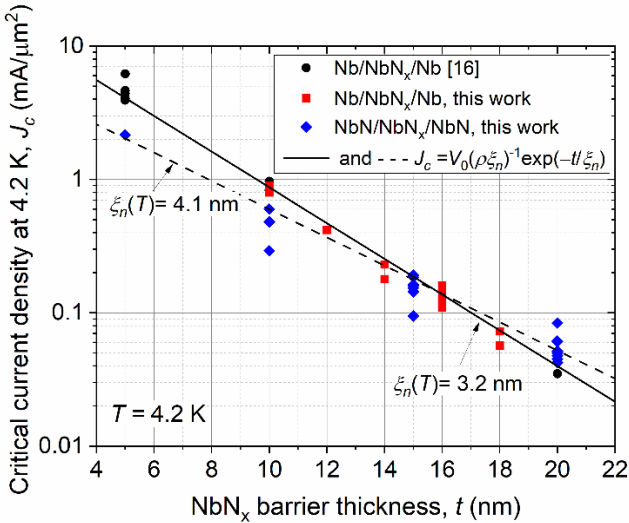

Fig. 5. Dependences of the Josephson critical current density at 4.2 K of Nb/NbN$_x$/Nb and NbN/NbN$_x$/NbN Josephson junctions on the thickness of the NbN$_x$ barriers. The data for Nb/NbNx/Nb junctions from [16], fabricated with the same NbN$_x$ deposition parameters, were also included. Solid and dashed lines are dependences (1). For Nb/NbN$_x$/Nb JJs (solid line), parameters $V_0$, $\xi_n$(4.2K) and the barrier resistivity values are identical to those determined in [16]: $V_0$ =4.4 mV, $\xi_n$(4.2K) = 3.2 nm, $\rho$ = 0.0071 Ω·cm. The dash line corresponds to $V_0$ =2.0 mV, $\xi_n$(4.2K) = 4.1 nm, and $\rho$ = 0.0071 Ω·cm, and seems to better describe NbN/NbN$_x$/NbN JJs.

CVCs and parameters of the fabricated Nb/NbN$_x$/Nb junctions were nearly identical to those presented in [16] and will not be repeated here.

*B. Scaling of $J_c$ and $I_cR_n$ with Barrier Thickness*

Dependences of the Josephson critical current density on the barrier thickness are shown in Fig. 5. For Nb/NbN$_x$/Nb junctions, we used identical deposition parameters to those used in [16]. The solid line is the dependence

$$J_c = \frac{V_0}{\rho \xi_n(T)} \exp\left(-\frac{t}{\xi_n(T)}\right) \quad (1)$$

following from the basic theory of SNS junctions [22], [23]. We used parameters $V_0$ =4.4 mV, $\xi_n$(4.2K) = 3.2 nm, the barrier resistivity $\rho = 0.0071$ Ω·cm, very similar to those previously determined in [16] for the NbN$_x$ barriers deposited at 20°C using N$_2$/(N$_2$+Ar)=0.905, the same as in this work. The data [16] obtained in 2022 and the new data agree very well indicating a good repeatability of the junction fabrication process.

The spread of the $j_c$ and junction resistance $R_nA$ values in Figs. 5,6 is caused by the variation of these parameters across the wafers as will be discussed separately.

Note also a factor of ~3x difference between the extracted resistivity of the NbN$_x$ barrier of about 7 mΩ cm and the in-plane resistivity of NbN$_x$ films in Table I and Fig. 1.

*C. Across-Wafer Variation of $J_c$*

Fig. 7 shows the typical variation of the Josephson critical current density of NbN/NbNx/NbN junction on 200-mm wafers. Twelve JJs with diameters from 0.7 μm to 2.2 μm were measured at each location specified in Fig. 7, and $j_c$ was determined from the slope of the linear dependence $I_c^{1/2} = (\frac{\pi j_c}{4})^{1/2} d$. The observed center-to-edge change in $j_c$ is about 34%. This change is caused by the center to edge variation of the barrier thickness which is seeing in the wafermaps in Fig. 1. Because of the very short coherence length in the barrier, even small changes in the deposited barrier thickness can cause large changes in the $j_c$. The observed change in $j_c$ in Fig. 7 requires only 1 nm change in $t$ from center-to-edge, as shown by the right axis in Fig. 7. The similar across-wafer radial variation of $j_c$ was also found for Nb/NbN$_x$/Nb junctions and shown in the bottom panel in Fig. 7.

The observed variation can be reduced by improving the barrier film thickness uniformity. This could possibly be done using a deposition system with wafer rotation and improved sputtering gas mixture distribution. For instance, for 40 nm Mo$_2$N films used as kinetic inductors [2], we were able to achieve the thickness standard deviation of $1\sigma = 2\%$ on 200-mm wafers, using a Connexion PVD tool from Veeco Instruments Inc. Nevertheless, a 2% thickness variation for the 10 nm barrier would result into more than 10% standard deviation of the $j_c$.

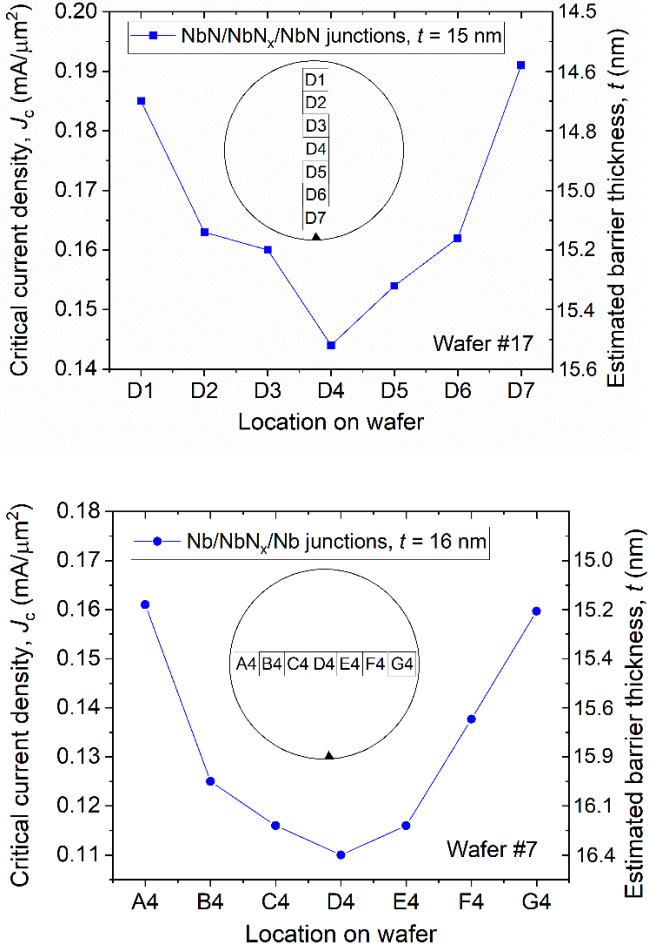

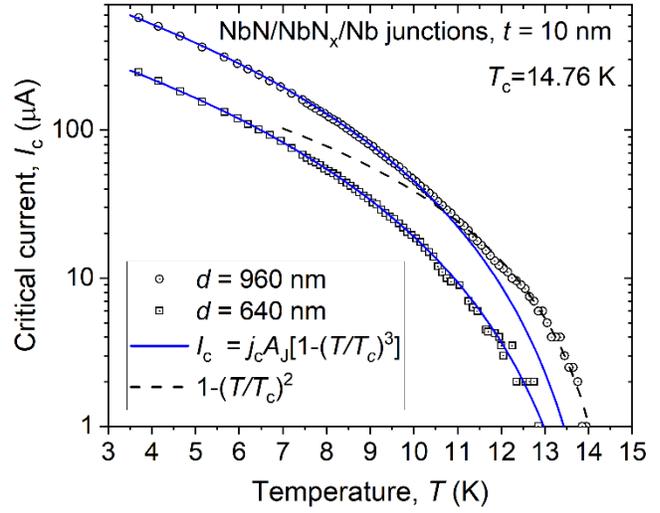

Fig. 7. The typical across-the-wafer variation of the critical current density of NbN/NbN$_x$/NbN (top panel, variation along the y-axis) and Nb/NbN$_x$/Nb (bottom panel, variation along the x-axis) Josephson junctions with reactively sputtered NbN$_x$ barriers with the nominal thickness of 15 nm and 16 nm, respectively. The right axis in both panels shows the estimated variation of the barrier thickness based on (1) with the coherence length extracted from the data in Fig. 5. The measurements were done on a junction test chip located on each 22 mm x 22 mm process control monitor (PCM) die corresponding to a 7x7 exposure matrix (A,B,…,G; 1,2,...,7) of a 248-nm Canon EX4 stepper used for the junction photolithography.

*D. Temperature Dependence of the Critical Current*

Yet another important characteristic for JJ applications in VLSI circuits is temperature dependence of the critical current, $I_c(T)$ in the temperature range of the circuit operation. Due to unavoidable heat generation during the circuit operation and poor thermal conductivity in multilayered stacks, temperature of the circuit or of its parts may rise decreasing $I_c$. To remain operational, the critical current change $\Delta I_c/I_c$ must be smaller than some value determined by the circuit design, typically less than 20 – 30%. Hence, using junctions with a weak $I_c(T)$ dependence near the operating temperature allows for a higher on-chip heat dissipation and larger integration scale, and gives larger operating margins than using JJs with a strong $I_c(T)$ dependence.

Fig. 8 shows the typical $I_c(T)$ dependence for NbN/NbN$_x$/NbN JJs with 10 nm barriers. In almost the entire temperature range from 4 K to ~ 13 K, the critical current strongly decreases with temperature as $I_c(T) \propto 1 - (T/T_c)^3$. Only very close to $T_c$, at $T/T_c \gtrsim 0.8$, the dependence becomes closer to $1 - (T/T_c)^2$ which is expected for the SNS and other proximity effect junctions with thick barriers, $t > \xi_N$ [22], [23], [24], [25]. Contrary to this strong dependence, $I_c$ of SIS junctions becomes almost temperature independent at $T \lesssim T_c/2$ [25], [26], [27]. For a comparison of the $I_c(T)$ dependence of Nb/NbN$_x$/Nb and Nb/Al-AlO$_x$/Nb junctions see [16].

Fig. 8. The typical temperature dependence of the critical current of NbN/NbN$_x$/NbN Josephson junctions with reactively sputtered NbN$_x$ barrier with the nominal thickness of 10 nm. Solid and dash curves show $1 - (T/T_c)^3$ and $1 - (T/T_c)^2$ dependences, respectively.

## V. CONCLUSION

In conclusion, we investigated NbN/NbN$_x$/NbN and Nb/NbN$_x$/Nb self-shunted Josephson junctions to evaluate whether they can potentially replace Nb/Al-AlO$_x$/Nb tunnel barrier junctions with high transparency barriers. We find that, unfortunately, all proximity-type junctions with deposited barriers demonstrate many features that make them inferior to the tunnel barrier junctions for applications in VLSI integrated circuits: larger nonuniformity of the properties across the wafers, lower $I_cR_n$ products, and stronger temperature dependence of the junction critical current.

Because of relatively low $I_cR_n$ products and a substantial $I_c(T)$ dependence, the best potential application of the proximity-type junctions with deposited barriers, in our view, is in adiabatic quantum flux parametron (AQFP) circuits [28]. AQFPs do not require high $I_cR_n$ products because they, anyway, need to operate at frequencies below ~7 GHz to be in the adiabatic regime [28]. Energy dissipation in AQFPs circuits is extremely low [29] and the circuit density is low as well [30], resulting in the extremely low heat power densities and making a potential temperature rise during the AQFP circuits operation the smallest possible among all other types of superconductor logics.

For the next round of NbN junctions development, we are planning to utilize the Connexion PVD tool providing about 2% barrier thickness nonuniformity. Unfortunately, we have not been able to get access to and evaluate commercial PVD, chemical vapor deposition (CVD), and atomic layer deposition (ALD) tools capable of achieving a variation in reactive growth of 1% or less. Theoretically, a custom system could be designed and built to alleviate the thickness variation constraint if necessary. One of the purposes of this work has been determining if such an investment of funds and labor is warranted based on the characteristics of the deposited-barrier junctions.

ACKNOWLEDGMENT

The STEM imaging of the junctions was done at the Center for Functional Nanomaterials at Brookhaven National Laboratory. We are very grateful to Kim Kisslinger for making the junction cross sections and taking the STEM images.

Any opinions, findings, conclusions, or recommendations expressed in this material are those of the authors and do not necessarily reflect the views of the Under Secretary of Defense for Research and Engineering or the U.S. Government. Notwithstanding any copyright notice, U.S. Government rights in this article are defined by DFARS 252.227-7013 or DFARS 252.227-7014 as detailed above. Use of this article other than as specifically authorized by the U.S. Government may violate any copyrights that exist in this article. The U.S. Government is authorized to reproduce and distribute reprints for Governmental purposes notwithstanding any copyright annotation thereon.